\documentclass[fleqn,usenatbib]{mnras}

\usepackage{mathptmx}
\usepackage{txfonts}

\usepackage[T1]{fontenc}
\usepackage{ae,aecompl}
\usepackage{amsmath}    
\usepackage{epsfig}

\newcommand{\co}{\mbox{$^{12}$CO}}
\newcommand{\coa}{\mbox{$^{13}$CO}}

\newcommand{\kms}{\mbox{km s$^{-1}$}}

\newcommand{\cc}{\mbox{cm$^{-3}$}}
\newcommand{\cmsq}{\mbox{cm$^{-2}$}}

\newcommand{\vlsr}{\mbox{$V_{{\rm LSR}}$}}

\title[Short title, max. 45 characters]{MNRAS \LaTeXe\ template -- title goes here}

\title[Striations in the Taurus molecular cloud]{Striations in the Taurus molecular cloud: Kelvin-Helmholtz instability or MHD waves?}
\author[]
{M.~Heyer$^1$, P.~F.~Goldsmith$^2$, U.~A.~Y{\i}ld{\i}z$^2$, R.~L.~Snell$^{1}$, E.~Falgarone$^3$, J.~L.~Pineda$^2$\\
$^{1}$Department of Astronomy, University of Massachusetts, Amherst, MA 01003, USA\\
$^{2}$Jet Propulsion Laboratory, 4800 Oak Groave Drive, Pasadena, CA 91109, USA\\
$^{3}$LERMA, CNRS UMR 8112, E\'cole Normale Supe\'rieure, 24 rue Lhomond, 75231 Paris Cedex 05, France
}

\date{Accepted XXX. Received YYY; in original form ZZZ}

\pubyear{2016}

\begin{document}
\label{firstpage}
\pagerange{\pageref{firstpage}--\pageref{lastpage}}
\maketitle


\begin{abstract}
The origin of striations aligned along the local magnetic field direction in the translucent 
envelope of the Taurus 
molecular cloud is examined with new observations of \co\ and \coa\ J=2-1 emission obtained with the 
10~m submillimeter telescope of the Arizona Radio Observatory.  These data identify a periodic 
pattern of excess blue and redshifted emission that is responsible for 
the striations. For both \co\ and \coa, spatial variations of the J=2-1 to J=1-0 line ratio are small 
and are not spatially correlated with the striation locations. 
A medium comprised of unresolved CO emitting substructures (cells) 
with a beam area filling factor less than unity at any velocity is 
required to explain the average line ratios and brightness temperatures.
We propose that the striations result 
from the modulation of velocities and the beam filling factor of the cells as a result of either 
the Kelvin-Helmholtz 
instability or 
magnetosonic waves propagating through 
the envelope of the Taurus  molecular cloud.  Both processes are likely common features in 
molecular clouds that are sub-Alfv\'enic 
and may explain low column density, cirrus-like features similarly aligned with the magnetic 
field observed throughout the interstellar medium in far-infrared surveys of dust emission. 
\end{abstract}
\begin{keywords}
ISM: magnetic fields -- ISM: kinematic and dynamics -- ISM: molecules -- MHD -- waves -- turbulence
\end{keywords}

\section{Introduction}                                               

Supersonic gas motion in Galactic molecular clouds is a key 
property that determines cloud evolution and the production of newborn stars. 
The observed gas motions are most generally attributed to turbulent flows in a 
medium with very large Reynolds number,
which reflects the ratio of inertial to viscous forces. The variation of velocity dispersions with spatial scales 
in molecular clouds strongly supports the presence of turbulent flows in these regions \citep{Larson:1981, Falgarone:1992, 
Heyer:2004}.  
Since the interstellar magnetic field threads molecular clouds, any description of gas motions and cloud structure 
must consider the role of the magnetic force. 

\citet{Arons:1975} attributed the supersonic line widths to hydromagnetic waves and derived three conditions for 
such waves to operate:
1) magnetic stress larger than the thermal gas pressure, 2) a weak field limit such that field fluctuations, 
${\delta}{\rm B}$ are much smaller than the static field, ${\rm B}_\circ$, and 3) wave period longer than 
the neutral-ion collision time that couples the neutral material to the magnetic field.  
Subsequent theoretical studies have further explored the role of hydromagnetic waves in molecular clouds
\citep{Langer:1978, Carlberg:1990, Mouschovias:1991, Gehman:1996, Basu:2010} with a recent 
comprehensive 
treatment provided by \citet{Mouschovias:2011}.  

Hydromagnetic instabilities can 
also imprint structure to the velocity and density fields of interstellar clouds.  
Specifically, 
the Kelvin-Helmholtz instability can develop from 
density inhomogeneities and 
velocity shear layers from turbulent or champagne-like flows.  The density perturbations generated 
by the instability 
may account for 
some of the complex structure found in molecular clouds \citep{Berne:2012, Hendrix:2015}.


A valuable target to examine the role of the magnetic field in cloud motions is the translucent 
envelope of the Taurus molecular cloud.  \co\ J=1-0 
imaging of the cloud identified a network of faint, narrow, elongated features, hereafter, called 
striations, that are aligned along 
the local magnetic field direction
\citep{Goldsmith:2008}. \citet{Heyer:2008} analyzed the \co\ and \coa\ J=1-0 emission from a subfield of the most prominent 
striations and 
found distinct velocity structure functions along axes parallel and perpendicular to the local field direction.   That is, 
gas velocities vary smoothly, if at all, along the magnetic field direction but exhibit higher spatial frequency differences 
transverse to the field. 
Such magnetically 
aligned velocity anisotropy is evident in MHD simulations only when the motions are sub-Alfv\'enic and the 
magnetic pressure is larger than the 
local thermal pressure \citep{Heyer:2008}.  The plane of the sky component of the magnetic field in 
this region of the Taurus cloud 
has been estimated by \citet{Chapman:2011} to be 12-37 $\mu$G based on the Chandrasekhar-Fermi 
effect \citep{Chandrasekhar:1953}.

In this study, we present new \co\ and \coa\ J=2-1 line emission data from the most prominent striations
in the Taurus cloud.  These observations have higher angular resolution and much higher signal to noise ratios than the CO J=1-0 
data presented by \citet{Goldsmith:2008} that surveyed much of the Taurus cloud complex.  The new data also enable an 
examination of the excitation conditions. In \S2, the collection of these new data is described. 
The results are presented in 
\S3, along with a summary of velocity variations and line excitation analysis. 
In \S4, we examine the Kelvin-Helmholtz instability and 
magnetosonic waves as the responsible agents for the appearance and kinematics of the striations.

\section{Data}
The Arizona Radio Observatory 10~meter Submillimeter Telescope on Mount Graham was used to observe the 
J=2-1 transitions of \co\ and \coa\ toward a subfield in the Taurus molecular cloud that 
exhibited bright filamentary features in the \co\ J=1-0 images presented by \citet{Goldsmith:2008}. 
Observations 
were made on
2014 November 28--December 1, 2015 January 28--29, and 2015 February 9--18.  
The frontend was the ALMA Band 6 prototype receiver, which has the capability for simultaneous dual polarization observations 
of \co\ 2--1  
in the Upper Side Band (USB) and \coa\ 2--1 in the Lower Side Band (LSB).  The spectrometers were filter 
banks with 
250~kHz resolution corresponding to 0.325~\kms\ and 0.341~\kms\ at 
230~GHz and 220~GHz respectively. 
The beam size at the \co\ 2--1 line line frequency is 32$\arcsec$. 

The observed field was centered 
on coordinates, RA(J2000)=04$^{\rm h}$48$^{\rm m}$ 48.29$^{\rm s}$, Dec(J2000)= 26$^\circ$ 39\arcmin\ 58.5\arcsec. 
The map was constructed from 14 subfields, each 10\arcmin$\times$10\arcmin\ in extent, using
the ``On-the-Fly'' (OTF) scanning method. 
To increase  the signal to noise ratio, 
each subfield was observed 2 to 4 times. 

Routines within GILDAS\footnote{http://www.iram.fr/IRAMFR/GILDAS}  were applied to 
convolve the irregularly sampled 
OTF data with a Gaussian kernel with 10\arcsec\ half-power beam width to yield a final Nyquist-sampled map.
The data were acquired on the ${\rm T}^{*}_{\rm A}$ antenna temperature scale and scaled up to main beam temperatures using 
the beam efficiency  of 0.74 for both J=2-1 transitions.
The achieved median 1$\sigma$ rms sensitivity in main beam temperature units for both \co\ and \coa\ is 0.11~K.

We also make use of the Taurus \co\ and \coa\ J=1-0 data collected by the FCRAO 14~meter telescope \citep{Narayanan:2008}.  These data have higher native spectral resolution but less
sensitivity as these are part of the larger survey of the Taurus molecular cloud. The J=1-0 data have also been corrected for the 
error beam pattern of the 14~m telescope so the antenna temperatures correspond to main beam temperature units.
To facilitate a comparison with the two data sets, all 
data were smoothed to an angular resolution (HPBW) of 50\arcsec\ and a velocity spacing of 0.325~\kms.  
The median 1$\sigma$ rms sensitivities of these 
smoothed data are 0.04~K and 0.03~K for \co\ and \coa\ J=2-1 and 
0.46~K and 0.24~K for \co\ and \coa\ J=1-0.

\section{Results}
\begin{figure*}
\begin{center}
\epsfxsize=18cm\epsfbox{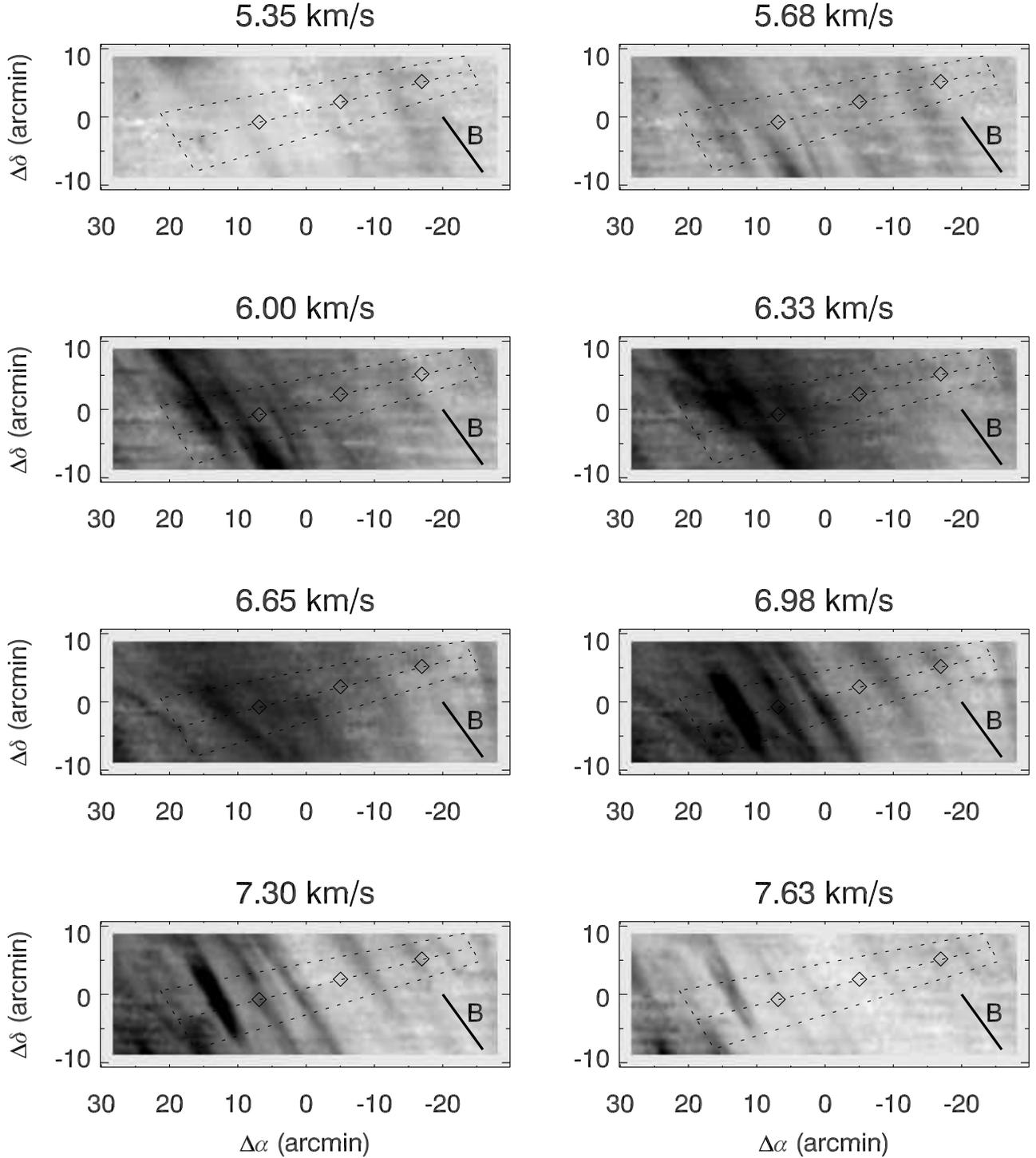}
\caption{Images of \co\ J=2-1 emission within separate spectroscopic channels.    
Halftones range 
from -0.2~K (white) to 2~K (black). The black line segment indicates the local magnetic field direction 
inferred from optical/IR polarization vectors.
The trapezoid shapes in each image show the length over which the spectrogram in Figure~\ref{fig3} are 
constructed and 
the width over which spectra are averaged.  
The diamond symbols along the central axis of the trapezoid denote lengths measured from the left 
edge of 0.5, 1, and 1.5~pc.  
Striations 
are evident in all channels but the first.
}
\label{fig1}
\end{center}
\end{figure*}

\begin{figure*}
\begin{center}
\epsfxsize=18cm\epsfbox{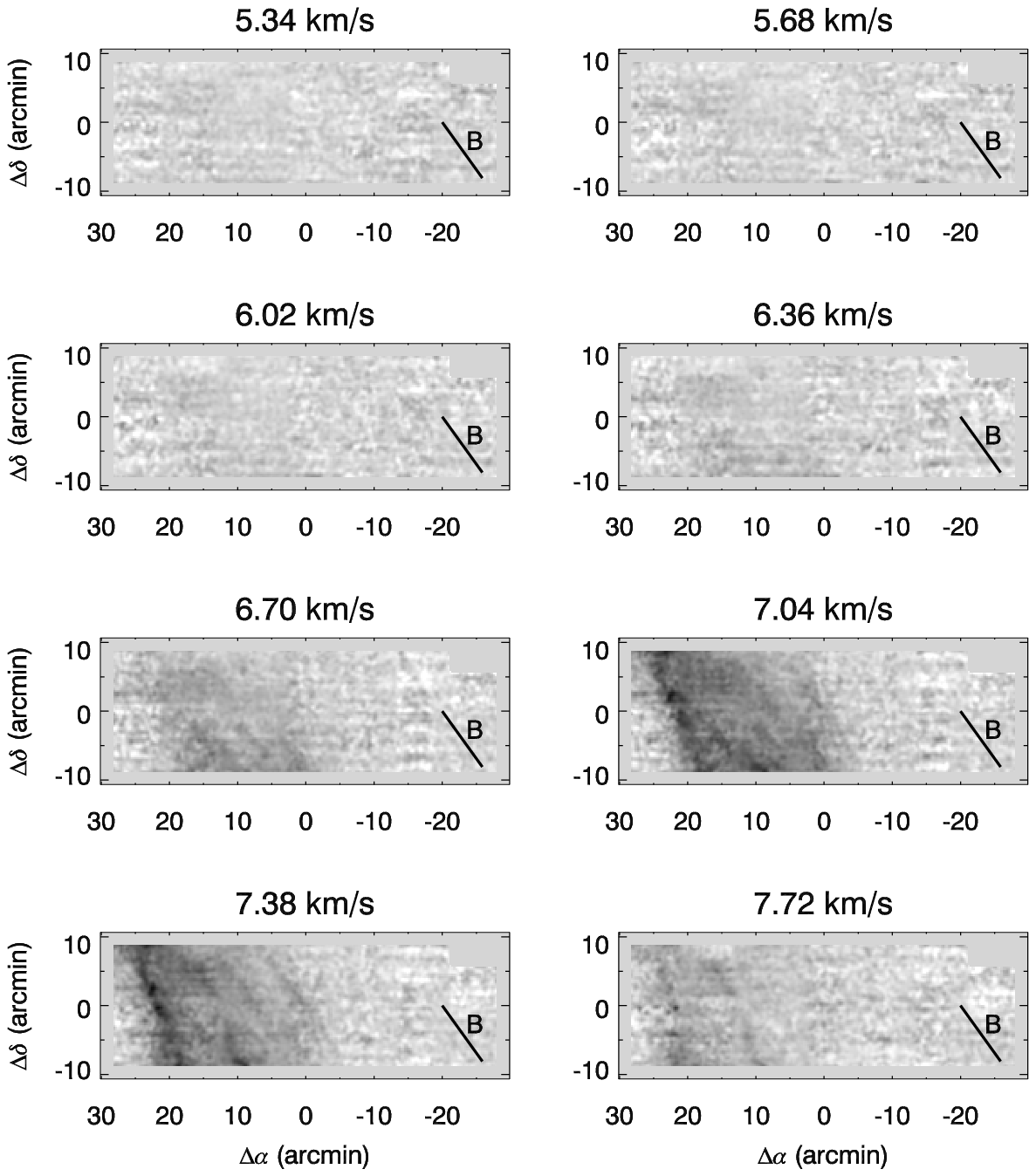}
\caption{Images of \coa\ J=2-1 emission within separate spectroscopic channels.   
Halftones range 
from -0.15~K (white) to 0.75~K (black). 
}
\label{fig2}
\end{center}
\end{figure*}

The \co\ J=1-0 data of the Taurus striations presented by \citet{Goldsmith:2008} demonstrate that these 
features are spatially coherent within velocity intervals less than 0.25~\kms.  
To effectively illustrate these features in the new data, we 
show the channel images of the \co\ and \coa\ J=2-1 emission in Figure~\ref{fig1} and Figure~\ref{fig2} respectively. 
In the \co\ images, the striations appear in each displayed velocity interval and are mostly parallel to each other.
The blue-shifted features are generally fainter and narrower 
than the red-shifted counterparts.   Striations are evident within the core velocity interval 
(6.3 $<$ \vlsr $<$ 6.7~\kms) but with less contrast with respect to brighter diffuse 
emission component. 
The \coa\ J=2-1 emission is very weak in this field.  The detected signal resides within the redshifted half of the line profiles and 
is mostly distributed within an extended component.  There are faint, elongated features that align with the 
brightest \co\ striations within the \vlsr=7.38 and 7.72 \kms\ images.  


Assuming spinning, elongated dust grains with minor axes aligned with the magnetic field, we calculate the local mean magnetic field direction from the set of 18 optical/IR polarization angles 
located within 1 degree of our field centre from the compilation by \citet{Heiles:2000}.  
The mean angle is 36$^\circ$ with a standard deviation of 7$^\circ$ indicating that the 
projected magnetic field is highly uniform in this sector of the Taurus cloud. This mean angle is 
comparable to the value of 27$^\circ$ derived from 353~GHz polarization of dust emission from 
the approximate center of our map 
\citep{Planck:2015}.
Figure~\ref{fig1} shows the alignment of the brightest striations along the local 
magnetic field direction inferred from the optical/IR polarization.

The spatial variation of the Taurus striations with velocity motivates an examination of the molecular line profiles within this field.  
Spectrograms (position-velocity images) are constructed from the \co\ and \coa\ J=2-1 and J=1-0 
data cubes along an axis running perpendicular to the striations, as illustrated by the trapezoid shown in each 
channel image of Figure~\ref{fig1}. 
The long axis of the trapezoid corresponds to the length of the spectrogram.  Spectra are 
averaged along the width of the trapezoid to gain signal to noise in the final spectrogram.   This width narrows along 
the length of the trapezoid due to the confines of the observed field so the resultant 
noise level is not constant along the spectrogram.
The spectral axes of each spectrogram are aligned to 
facilitate the calculation of isotopic and line ratios.    The rms noise levels per channel are 0.014~K, 0.012~K, 0.14~K, and 0.066~K
 for the \co\ J=2-1, 
\coa\ J=2-1, \co\ J=1-0 and \coa\ J=1-0 spectrograms respectively.   The spectrogram lengths for each data set are 
converted into parsecs assuming a distance to the Taurus molecular cloud of 140~pc.

\begin{figure*}
\begin{center}
\epsfxsize=18cm\epsfbox{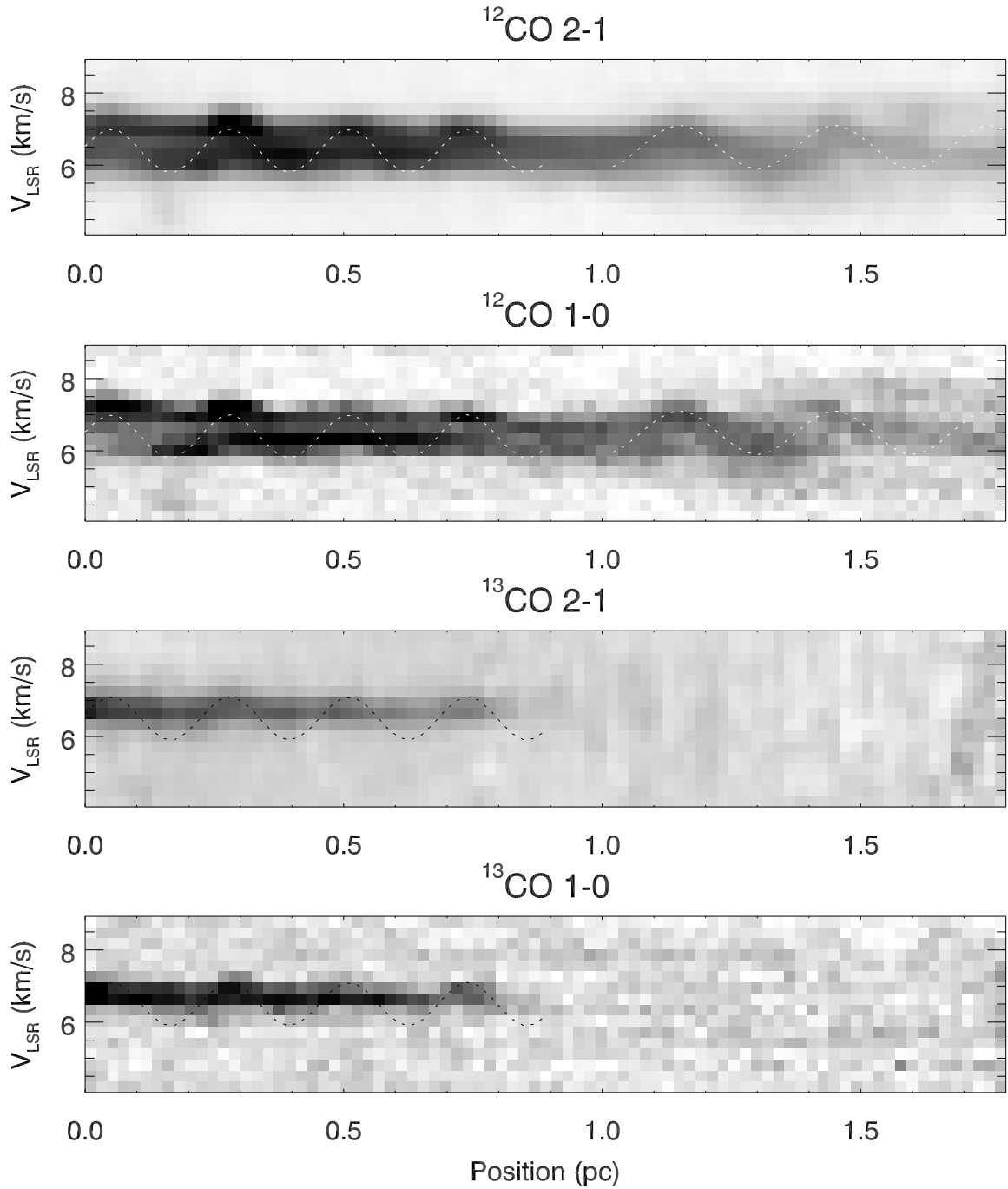}
\caption{Spectrograms along an axis perpendicular to the striations and local magnetic field direction 
for (top to bottom) \co\ J=2-1, \co\ J=1-0, \coa\ J=2-1, and \coa\ J=1-0.  The corresponding halftone ranges are:
-0.1~K (white) to 2.0~K (black); -0.2~K (white) to 3.0~K (black); -0.1~K (white) to 0.5~K (black); 
-0.15~K (white) to 1.0~K (black).  The dotted lines are ``by-eye'' fits of sine waves to oscillating excess emission 
in the blue and redshifted shoulders of the line profiles. 
These velocity oscillations lead to the
interleaving of the blue and redshifted striations. 
} 
\label{fig3}
\end{center}
\end{figure*}

The spectrograms for each transition and isotopologue are shown in Figure~\ref{fig3}.  
There is a clear periodic pattern in the \co\ J=2-1 and J=1-0 spectra, which is not detected 
in either \coa\ spectrograms.  
Excess emission oscillates between the blue and red shoulders of the line profile along the spectrogram axis.  
This effect is also seen in the  variation of the velocity centroid and skewness of the spectrogram line profiles. 
The overlayed sine waves, guided by the \co\ J=2-1 data, are drawn to 
illustrate these velocity oscillations.  For offset positions less than 0.9~pc, the data are well-described by a single 
sine wave with a projected wavelength of 0.23~pc.  
For the fainter 
features with offset positions greater than 0.9~pc, the velocity pattern shifts in phase and increases in wavelength 
to 
0.3~pc. 
These quasi-periodic velocity oscillations lead to the
interleaving of the brightest 
blue-shifted striations (5.35 $<$ \vlsr\ $<$ 6~\kms) 
with the brightest, red-shifted striations
(7.0 $<$ \vlsr\ $<$ 7.6~\kms) shown in Figure~\ref{fig1}. 

\begin{figure*}
\begin{center}
\epsfxsize=18cm\epsfbox{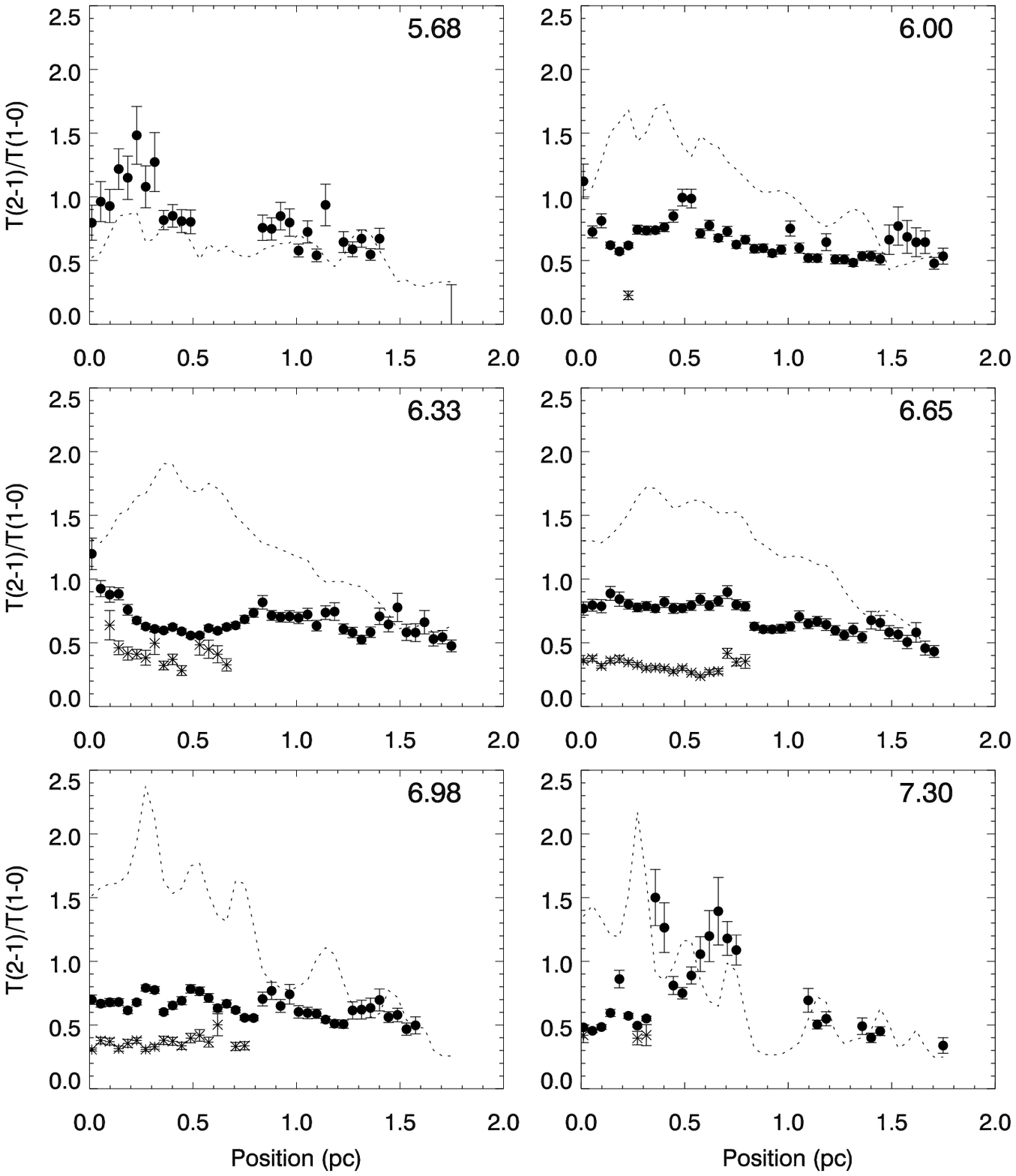}
\caption{Variation of antenna temperature ratios (${\rm T}_{\rm mb}$(2-1)/${\rm T}_{\rm mb}$(1-0)), for \co\ (solid circles) and \coa\ (asterisk) along the 
spectrogram axis. 
The dotted line corresponds to the \co\ J=2-1 brightness temperature for each velocity channel.  
There is 
no correspondance between positions of enhanced excitation and the location of the striations.
} 
\label{fig4}
\end{center}
\end{figure*}

Establishing the physical conditions of the gas within and between the striations for a given velocity interval 
can help guide a description 
of the physical origin of these features. Do the striations correspond to sites of enhanced volume density,  
warmer gas temperatures, and larger CO abundances relative to the background 
as may be expected for a shock, instability, or longitudinal MHD wave?
Ideally, the set of observed CO lines offers limited constraints on the physical conditions within 
the observed field.  Since the spatial variations of the emission are a strong function of 
velocity, we first examine the J=2-1 to J=1-0 
line ratios in individual velocity channels along the 
spatial extent of the \co\ and \coa\ spectrograms as shown in 
Figure~\ref{fig4}.  
The rms errors for each transition are propagated to derive an uncertainty in the ratio. 
For clarity, only ratios with signal to noise greater than 5 are displayed.  The trace of the J=2-1 antenna 
temperature for each velocity interval, ${\rm T}_{\rm mb}{\rm (v)}$, is also drawn to place the line ratio in 
context with the striations. 
Figure~\ref{fig4} illustrates that there is detected \co\ J=2-1 emission at all positions along the spectrogram lengths.
The striations are enhancements of signal relative to this background component.  The fractional increments of the 
striations relative to the background signal varies with position and velocity -- 
ranging from 10 to 100\% with the strongest boost in the \vlsr=6.98 and 7.30 \kms\ channels.
Overall, there is no clear relationship between the measured line ratios and striation locations.   
There are both small ratio increments and decrements at the striation positions and within the gaps between the 
striations.
The \co\ line ratios profiles are flat within the core of the line (6.33 and 6.65 \kms\ channels).  
The ratio derived 
from the lower opacity \coa\ lines 
should be more sensitive to local density variations. Yet, the \coa\ ratios are also flat along the 
length where the ratios are reliably measured.  

While the observed 2-1/1-0 line ratios are evidently not sensitive enough to measure variations in temperature and density 
within the narrow velocity range of the striations, the range and mean 
values do provide a coarse measure of the average physical conditions in the field.
Average spectra for each isotopologue and transition are constructed from the data cubes using equal weighting.
Integrated intensities
are calculated from these average spectra over the velocity interval 4.5 to 8 \kms.  The corresponding 
2-1/1-0 line ratios are 0.68 and 0.34 for 
\co\ and \coa\ respectively with propagated random errors less than 1\%. 
Gaussian profiles are fit to each average spectrum to determine the peak intensity and 
the full width at half maximum line width, ${\delta}{\rm v}$ that are required for non-LTE modeling. 
Peak antenna temperatures are 1.3~K and 1.9~K for the \co\ J=2-1
and J=1-0 lines respectively;  and 0.14~K and 0.45~K for the \coa\ J=2-1 and J=1-0 transitions. 
The deconvolved, one dimensional, full width half maximum (FWHM) line widths are 1.7~\kms\ for the \co\ 
lines and 0.7~\kms\ for the \coa\ lines.

The line radiative transfer and excitation model,
RADEX \citep{vanderTak:2007}, is used to compute model
line ratios and intensities for a set of cloud conditions
(temperature, density, column density). With two transitions
and two isotopologues, we might hope to identify
a simple set of cloud conditions for which the non-LTE model
matches the measured line ratios and brightness
temperatures.  However this is not the case.
The primary impediment to a self-consistent model
is the large line ratio and low brightness temperature
of \co, and simultaneously, low \coa\ line
ratio. 

A simple argument can be made for unresolved, inhomogeneous
density structure in the Taurus envelope that can explain
the observed line ratios.
There is a family
of model cloud conditions that are consistent with
the line ratio of the optically thin \coa\ lines. For example,
models with kinetic temperature and volume density, 
(T$_{\rm k}$,n)=(10~K,880~\cc), (20~K,330~\cc), and  (30~K,210~\cc), 
all match the
mean \coa\ 2-1/1-0 line ratio of 0.34. Using any of these
density and temperature combinations does not produce the
larger observed \co\ 2-1/1-0 line ratio of 0.68 unless the optical
depths of the \co\ lines are sufficiently large
for radiative trapping to increase the excitation of \co\ relative
to \coa. Although \co\ is subthermally excited, any model
that matches the observed \co\ line ratio predicts line
intensities $\sim$5 times larger than observed.
This difference between the modeled and observed line intensities can
be removed if much of the CO emission originates in a medium
composed of unresolved substructures, hereafter cells,  
that fill a fraction ($\sim$20\%) of
the antenna beam area at any velocity.  
Such small scale density or abundance 
inhomogeneities in molecular clouds have been inferred by previous 
studies based on similar excitation arguments
as those described here 
\citep{Goldsmith:1975, Snell:1984,Tauber:1991,Falgarone:1996,Falgarone:1998,Hily-Blant:2007}.
\citet{Falgarone:1996} derive 
an upper limit to the size of cells in the envelope 
of the Perseus-Auriga cloud to be 34~AU based on the smoothness of 
line profiles observed with high signal to noise and the requirement of moderate
beam averaged optical depths.  
With only 2 CO isotopologues, 2 rotational transitions, and moderate signal to noise, 
we can not establish any more
precisely the conditions, sizes, or the number of cells in this part of the Taurus cloud. 

The clumpy medium description of the cloud envelope gas 
implied by the modeling of the line ratios and intensities offers an alternative 
view of the  
striation molecular emission. 
The variation of \co\ surface brightness across the spectrogram axis 
results from the spatial modulation of the beam filling factor of unresolved cells at a given velocity 
rather than changes of the 
excitation conditions.  The observed invariance of the line ratios reflects 
the uniformity  
of temperature and density within the cells.  The physical mechanisms responsible for this 
modulation of filling factor and velocity are discussed in \S4.

\section{Discussion}

The data described in the previous section and the results from earlier studies point to the magnetic field 
as the responsible agent for the aligned striations in the envelope of the Taurus molecular cloud. 
The improved signal to noise of the J=2-1 data, relative to the J=1-0 data,
identifies a quasi-periodic pattern of \co\ emission that switches between the blue and red shoulders of the 
observed line profiles as shown in Figure~\ref{fig3}.  This oscillatory velocity behavior produces the 
interleaving of the blue and redshifted striations with each other.  The other key characteristic of the striations is the 
anisotropy of velocities in this region \citep{Heyer:2008, Heyer:2012}.  Velocity profile shapes and centroid velocities 
vary smoothly, if at all, along the 
length of the striations and magnetic field. Perpendicular to these directions, the molecular gas velocities 
exhibit higher spatial frequency variations, as demonstrated in Figure~\ref{fig3}. 
This magnetically aligned velocity anisotropy is only possible in a medium for which the 
gas motions are sub-Alfv\'enic \citep{Heyer:2008, Esquivel:2011}.  
Modeling 
of the line ratios and intensities of the J=2-1 to J=1-0 transitions suggests the medium is inhomogenous
and that the 
CO emission from the striations is enhanced above the background due to an increased filling factor of 
small, unresolved condensations   
within the beam.

The \co\ striations are unlikely due to hydrodynamic shocks. First, there is no signature of a strong 
density enhancement that would be expected from an isothermal or adiabatic shock.  
Given the orientation of the striations, a shock would be 
produced by flows moving perpendicular to the magnetic field.  Yet, there is no curvature in structure of the magnetic field. 
Such flow would have to be broadly distributed and spatially coherent to 
produce such a planar shock front.  
Finally, such shocks would not produce the interleaved 
pattern of blue and redshifted features.  

\subsection{Kelvin-Helmholtz Instability}
The regular sequence of the brightest striations 
could be the product of the Kelvin-Helmholtz (K-H) instability that arises along the interface of two contiguous 
layers with different tangential velocities and volume densities.
In the Taurus envelope, such layers could be two contiguous, turbulent eddies 
or the interface between warm, neutral
atomic gas and cold, neutral material.
The K-H instability produces alternating zones of high and low
pressure that correspond to sites of higher and lower gas
density.  For a clumpy medium, this could be reflected in the periodic variation of the 
beam filling fraction of dense cells inferred from the line ratios.  

\citet{Berne:2012} and \citet{Hendrix:2015} examined 
the role of the magnetic K-H instability in producing the ripples of dust emission observed near the interface of the HII region 
and molecular cloud in Orion \citep{Berne:2010}.  
In their model, the interface is in the x-z plane, the velocity and density gradients are along the y-axis, and 
the magnetic field is 
at angle $\theta$ to the z-axis.
The spacing of the Orion ripples falls within the range of 
instability wavelengths (0.06-0.6~pc for the conditions in Orion) 
but only if $|\theta| <$25$^\circ$. 
Viewed from a line of sight along the y-axis (viewing the x-z plane), one 
would observe periodically spaced ripples along the x-axis that stretch vertically along the magnetic field direction (z-axis). 
Magnetic fields oriented transverse to the flow direction do not impact the wavelength or growth rate of the K-H intability
\citep{Chandra:1961}.

The environment of the Taurus striations is distinct from that of the region of the Orion ripples.  In Orion, the two layers 
are hot, ionized, low density gas and cold, dense molecular gas respectively,
 and the flow velocities are $\sim$10~\kms. In the Taurus envelope, 
the density and velocity differences are small, based on the limited variation of the CO 2-1/1-0 line ratios and the 
observed velocity dispersion ($\sim$0.7~\kms).  For two layers  with densities $\rho_1$ and $\rho_2$, and velocities, ${\rm u}_1$ and ${\rm u}_2$, 
the maximum allowed wavelength of the K-H instability limited by gravity is
\begin{equation}
\lambda_{\rm KH,max}=\frac{2\pi}{\rm g} \frac{\alpha_1\alpha_2}{\alpha_1-\alpha_2} (u_1-u_2)^2 ,
\end{equation}
where $\alpha_1=\rho_1/(\rho_1+\rho_2)$, $\alpha_2=\rho_2/(\rho_1+\rho_2)$,
 and g is the acceleration due to gravity 
\citep{Chandra:1961}. 
Normalizing to the Taurus envelope conditions
\begin{equation}
\lambda_{\rm KH,max}=6.1 \left( \frac{2{\times}10^{21}\; {\rm cm}^{-2}} { {\rm N}_{\rm H}} \right) \left(\frac{\alpha_1(1-\alpha_1)}{2\alpha_1-1}\right) \left(\frac{({\rm u}_1-{\rm u}_2)}{0.7\; {\rm kms}^{-1}}\right)^2 \; \;{\rm pc}, 
\end{equation}
where we have taken g=$\pi {\rm G} \mu {\rm m}_{\rm H} {\rm N_H}$.  For these conditions, the observed striations are well within this wavelength upper limit 
unless $\alpha_1$ approaches 
unity.  
For instabilities developing at the interface between layers of warm (WNM) and cold neutral (CNM) gas, the 
density contrast can be much higher.  Taking fiducial volume density values  of 0.5~\cc\ and 50~\cc\ for the WNM 
and CNM respectively and 
a velocity 
difference between the layers of $\sim$3~\kms, then $\lambda_{\rm KH,max}$ reduces to $\sim$ 1~pc. 


The K-H instability supports the appearance of multiple wavelengths, as observed in the target field in Taurus as long as 
$\lambda < \lambda_{\rm KH,max}$.
Using 3D simulations of the magnetic K-H instability, 
\citet{Matsumoto:2007} found that the local 
vortices become turbulent at later times. This non-linear behavior
causes these regions to fragment into even narrower 
features and could qualitatively describe the very small
spacing between the striations that are evident in the \vlsr=5.68 \kms\
image of Figure~\ref{fig1}.

While the K-H instability can replicate the general appearance of the Taurus striations,
(see Figure~8 in \citealt{Hendrix:2015}), two questions must be considered.
First, what is the Alfv\'enic Mach number of the differential flows required for the instability?
For gas flows with a component perpendicular to the magnetic field that is super-Alfv\'enic, 
one would expect large scale distortions of the 
field geometry.
Yet, the magnetic field in this part of the Taurus cloud appears 
uniform over angular scales 
of 1-2 degrees \citep{Planck:2015}. Thus, any flows responsible for triggering the K-H instability 
in Taurus 
must be sub-Alfv\'enic.
Second, can
the K-H instability 
produce the periodic velocity pattern 
illustrated in Figure~\ref{fig3}?  In the 3D MHD simulations of
\citet{Matsumoto:2007} with the ratio of gas pressure to magnetic pressure equal to 0.1, 
the velocity field shows converging 
streamlines towards the high pressure zones and more
circulatory motions around the low pressure vortex.  The converging flows 
must be asymmetric in density or temperature to generate a velocity displacement to 
lower or higher velocities.  Such asymmetry must vary periodically 
to produce the spatially interleaved blue and red-shifted striations while also 
be coherent along the striations to account for the observed anisotropy. 

\subsection{MHD Waves}
The concept of gas motions in molecular clouds being a result of propagating hydromagnetic waves was introduced 
by \citet{Arons:1975}. 
\citet{Mouschovias:2011} derive the dispersion relationships and describe the allowed modes for MHD waves propagating 
parallel and perpendicular 
to the local magnetic field direction as well as the more general case of an arbitrary angle.  The 
two primary modes relevant to this study are neutral transverse Alfv\'en and neutral 
magnetosonic waves.  The orientation of the striations relative to the local magnetic field, the periodic velocity 
structure perpendicular to the field and the absence of velocity structure along the field excludes the 
neutral, transverse, Alfv\'en wave as the cause of the striations.  
Such transverse disturbances produce velocity fluctuations along the 
striations, while also distorting the local field direction.  

If MHD waves are responsible for the striations, then these must be longitudinal magnetosonic 
waves.  As such waves propagate perpendicular to the mean magnetic field direction, 
the gas and magnetic field are alternately 
compressed and rarified.  In the case of a clumpy medium, the wave disturbance modulates the 
beam filling factor as oscillating thermal and magnetic pressure displace the cells. 
The effect is to 
vary the number of cells within a velocity interval over the solid angle of the telescope beam or 
resolution element.
The striations appear over lengths at which the velocities are  
coherent in response to the longitudinal disturbance.   This wave-induced spatial coherence of velocity 
along a given striation 
naturally accounts for magnetically aligned velocity anisotropy that is expressed as different indices of 
the velocity structure functions calculated parallel and perpendicular to the 
local field \citep{Heyer:2008,Heyer:2012}.  

MHD waves propagating within molecular clouds are expected to have a limited range of wavelengths owing to neutral-ion gas 
coupling and gravity \citep{Mouschovias:2011}.  An upper limit to the wavelength is 
set by the magnetic Jeans' length, $\lambda_{\rm J,mag}$, 
\begin{equation}
\lambda_{\rm J,mag}=2\pi {\rm v}_{\rm ms,n} \tau_{\rm ff} ,
\end{equation}
where ${\rm v}_{\rm ms,n}=({\rm v}_{\rm A,n}^2+{\rm c}_{\rm n}^2)^{1/2}$ is the magnetosonic speed for neutral gas, 
${\rm v}_{\rm A,n}={\rm B}/(4{\pi}\rho)^{1/2}$
is the 
Alfv\'en velocity for neutral gas with mass density $\rho$, ${\rm c}_{\rm n}=0.27 ({\rm T}/20~{\rm K})^{1/2}$~\kms\ is the thermal 
sound speed, and $\tau_{\rm ff}=(4{\pi}{\rm G}\rho)^{-1/2} = 0.8 (500~{\rm cm}^{-3}/{\rm n_H})^{1/2}$~Myr is the free-fall time.
This results in 
\begin{multline}
\lambda_{\rm J,mag}=4.8 \left( \frac{ {\rm B} } { 15~{\mu}{\rm G} } \right) \left( \frac{ 500~{\rm cm}^{-3} }{ {\rm n_H} } \right) \\
        \times \left[ (1 + 0.08 \left( \frac{ {\rm T} }{ 20~{\rm K} } \right) \left( \frac{ 15~{\mu }{\rm G} } { {\rm B} }\right)^2 \left(\frac{{\rm n}}{500~{\rm cm}^{-3}}\right)\right]^{1/2} \;\; {\rm pc}.
\end{multline}
For disturbances with wavelengths larger than $\lambda_{\rm J,mag}$, the gravitational force is larger than the magnetic force
and the localized region collapses \citep{Mouschovias:1991}. 

A mininum wavelength is imposed by the need for frequent neutral-ion collisions to transfer the magnetic force 
to the bulk of the material \citep{Kulsrud:1969, Arons:1975}.  From the dispersion relations derived by \citet{Mouschovias:2011}, 
the minimum wavelength for a neutral, magnetosonic wave is 
\begin{equation}
\lambda_{\rm ms,n} = \lambda_{\rm A,n} \frac{{\rm v}_{\rm A,n}}{{\rm v}_{\rm ms,n}} , 
\end{equation}
where $\lambda_{\rm A,n}$ is the Alfv\'en length \citep{Mouschovias:1991}.  For
an electron fraction, $x_{\rm i}$,
\begin{equation}
\lambda_{\rm A,n} = \pi {\rm v}_{\rm A,n}\tau_{\rm ni} = 0.02 \left(\frac{\rm B}{15~{\mu}{\rm G}}\right) \left(\frac{500~{\rm cm}^{-3}}{{\rm n_H}}\right)^{3/2} \left(\frac{10^{-5}}{x_{\rm i}}\right) \;\; {\rm pc},
\end{equation}
where $\tau_{\rm ni}$ is the neutral-ion collision time.
For magnetosonic disturbances with wavelengths less than $\lambda_{\rm ms,n}$, the magnetic field diffuses through the 
neutrals on time scales shorter than the 
neutral-ion collision time  leaving the neutral particles uncoupled to the magnetic field.   
In this regime, the wave is rapidly damped \citep{Mouschovias:1991}.

The column density, ${\rm N_H}$, in the observed field is low (1-2$\times$10$^{21}$ \cmsq) based on infrared-derived extinction and CO 
column densities \citep{Pineda:2010}. In this regime, both far UV radiation and cosmic rays contribute to the gas ionization. 
Using the Meudon PDR Model \citep{LePetit:2006}, we have calculated the ionization fraction with the following primary model parameters: 
radiation field of 1 Mathis field illuminated from both sides of the slab, a hydrogen density of 500~\cc,
a maximum depth of 3~$A_{\rm v}$, and a cosmic ray ionization rate of 5$\times$10$^{-17}$
s$^{-1}$. The resultant electron fraction at a depth of 2 magnitudes,
measured from the front face of the cloud, is 7.8$\times$10$^{-5}$.  
In this low volume density, diffuse environment, the minimum wavelength could be as small as 0.01~pc.

The projected wavelength for the brightest, and most repetitive striations in 
the \vlsr=6.98~\kms\ channel is 0.23~pc. As the striations are 
likely in a plane inclined to the plane of the sky by angle, $\theta_{\rm i}$, the actual wavelength is 0.23/cos($\theta_{\rm i}$) pc. 
Even accounting for these projections, 
the striations satisfy the wavelength conditions, $\lambda_{\rm ms,n} < \lambda < \lambda_{\rm J,mag}$.

While the observed wavelengths comply with the restrictions imposed by gravity and neutral-ion collision times, the 
magnetosonic waves do not fully account for the observations.  For a sinusoidal variation of gas spatial displacement,
${\rm s}={\rm s}_{\rm max}{\rm cos}({\rm kx}-{\omega}{\rm t})$, caused 
by the longitudinal wave disturbance, the velocity and density (beam filling factor) 
vary as $\partial {\rm s}/\partial {\rm t}$ and $\partial {\rm s}/\partial {\rm x}$ respectively, 
where ${\rm s}_{\rm max}$ is the maximum displacement, ${\rm k}=2\pi/\lambda$ and $\omega$ is the angular frequency of the wave. 
Density or beam filling fraction and velocity maxima, which may be related to the striations, occur when 
${\rm kx}-{\omega}{\rm t}={\rm n}{\pi}$ and {\rm n} is an even integer, 
while 
density and velocity minima occur when {\rm n} is an odd integer. 
Depending on the direction of the wave propagation (sign of k) 
and the angle of propagation with respect to the plane of the sky (since we only measure the radial component of the velocity 
shift), one expects a correspondance between the maximum velocity and maximum density (beam filling factor). 
 Similarly, there should be 
minima in the density (beam filling factor) 
 where the velocity is minimum.  Such a condition is only partially observed as the striations occur at both 
blue and red shifted velocities.  Evidently, interstellar 
gas flows  are more complicated than the idealized, single wave disturbance discussed here. 

\subsection{MHD Waves and K-H Instabilities in Molecular Clouds}
With the current data, we can not definitely assign either the K-H instability or 
magnetosonic waves  as 
the physical origin of the  Taurus striations. 
Both processes qualitatively produce such features aligned along the local magnetic field direction 
and predict density enhancements or increased beam filling fractions at the positions of the striations.  
The observed projected wavelengths of the striations are smaller than the wavelength upper limits set by gravity 
for both the K-H instability and wave disturbance
and larger than the minimum wavelength imposed by neutral-ion collision frequency. 
The oscillatory behavior of the line profiles can be partially reproduced by a magnetosonic wave propagating transverse to the local 
field direction but inclined to the plane of the sky.  Compression and rarefaction of the gas, represented by 
beam filling fraction, corresponds to the 
peak and trough of the wave depending on the direction and inclination of the wave propagation.   The wave model also 
describes the magnetically aligned velocity anisotropy.

The striations are not isolated to this small subregion within the Taurus cloud.  Wispy structures with similar velocity anisotropy 
aligned along the local field direction 
are found throughout the low column density regime of the Taurus cloud \citep{Heyer:2012}.  
The identification of 
these striations within the cloud envelope 
likely results from  a favorable viewing angle, 
a well-resolved wavelength of a single disturbance, and
a relatively limited depth into the 
cloud.  In the high column density regions of the cloud, corresponding to higher surface brightness CO emission, such striations 
are not 
apparent due to several factors.  In this regime, such features are masked by brighter emission from other structures along the line of 
sight and the beam area filling fraction within a small velocity interval approaches unity offering little or no contrast.  
For the K-H instability, the maximum allowed wavelength would be smaller owing to increased column density and reduced 
turbulent velocity differences at smaller scales. Similarly, only smaller wavelength magnetosonic modes are allowed 
in regions of high volume density and reduced ionization.  At some point, such features are not resolved by the
moderate angular resolution of these data. 
Finally, waves propagating from different directions can interact with each other in the central regions of the cloud.
Such interactions can distort any simple, recognizable pattern of an isolated wave. 

In a more general context, the presence of magnetosonic waves and K-H instabilities, revealed by striations, 
 should be common 
mechanisms in any sub-Alfv\'enic molecular cloud threaded by the interstellar magnetic field with sufficient ionization to 
couple the neutral gas to ions. 
Indeed, striations are observed throughout the cold, neutral ISM in the dust emission maps of IRAS, {\it Herschel}, and {\it Planck } missions 
and imaging of the HI~21cm line \citep{Clark:2014}.  Many of these low column density 
filamentary features are aligned with the local magnetic field direction \citep{Planck:2015}. 
The ubiquity of striations throughout the cold, neutral ISM, generated by K-H instabilities and/or  
magnetosonic waves illustrates an important role of the interstellar magnetic field in modulating gas 
motions.

\section{Conclusions}
\co\ and \coa\ J=2-1 data collected by the ARO 10~m telescope are analyzed in conjunction with available J=1-0 data to 
investigate the nature of striations that are aligned along the magnetic field in the Taurus molecular cloud. 
The high sensitivity of the J=2-1 data identify spatially oscillating blue and red shoulder line profile components corresponding 
to the striations.  
A medium 
comprised of unresolved cells that are responsible for the CO emission is inferred from the 
\co\ and \coa\ 2-1/1-0 line ratios and intensities. 
The striations result from the 
modulation of the velocities and the beam area filling faction of the cells by either a Kelvin-Helmholtz 
instability or magnetosonic waves propagating through the 
envelope of the Taurus cloud.  Such processes may explain the appearance of similar striations observed 
throughout the cold, neutral interstellar medium in images of dust and gas emission. 

\section*{Acknowledgments}
The authors thank L. Ziurys for granting additional time at the Arizona Radio Observatory to improve 
the sensitivity of the data and J. Bieging for guidance on calibration of the data. 
The Heinrich Hertz Submillimeter Telescope is operated by the Arizona Radio Observatory, 
which is part of Steward Observatory at The University of Arizona. 
The Arizona Radio Observatory is funded 
in part by National Science Foundation grant AST-1140030 to The University of Arizona.

\bibliographystyle{mnras}
\bibliography{cite_striae} 

\end{document}